\documentstyle[12pt]{article}
\input{psfig}
\begin{document}

\author{{\bf J. C. Anjos} \and {\bf G. Herrera} \thanks{
Permanent address: Centro de Investigaci\'{o}n y de Estudios 
Avanzados, Apdo. Postal 14 740, M\'{e}xico 07000, DF, Mexico.
e-mail: gherrera@fis.cinvestav.mx} \and 
{\bf J. Magnin} \thanks{e-mail: jmagnin@lafex.cbpf.br} \and 
{\bf F.R.A. Sim\~{a}o} \\
{\small Centro Brasileiro de Pesquisas F\'{\i}sicas} \\
{\small Rua Dr. Xavier Sigaud 150}\\
{\small 22290-180, Rio de Janeiro, Brazil}}

\title{Production and polarization of the
$\Lambda _c^{+}$ and the charm of the proton
\thanks{This work was partially supported by Centro Latino 
Americano de F\'{\i}sica (CLAF).} }

\date{}
\maketitle

\begin{tabular}{rrl}
PACS & 13.60.Rj & (Baryon production)\\
     & 13.87.Fh & (Fragmentation into hadrons)\\
     & 14.20.Lq & (Charmed baryons)
\end{tabular}

\begin{abstract}
We propose a two-component model involving the parton fusion mechanism 
and recombination of a $ud$ valence diquark with a sea $c$-quark 
of the incident proton to describe $\Lambda_c^+$ inclusive production 
in $pp$ collisions.
We also study the polarization of the produced $\Lambda_c^+$ 
in the framework of the Thomas Precession Model for polarization. 
We show that  a measurement of the $\Lambda_c$ polarization is a sensitive 
test of its production mechanism. In particular the intrinsic charm
model predicts a positive polarization for the $\Lambda_c$ 
within the framework of the Thomas Precession Model,
while according to the model presented here
the $\Lambda_c$ polarization should be negative.
The measurement of the $\Lambda_c$ polarization provides a close 
examination  of intrinsic charm Fock states in the proton and gives 
interesting information about the hadroproduction of charm.
\end{abstract}

\newpage

\section{Introduction}

The production mechanism of hadrons containing heavy quarks is not well
understood. Although the fusion reactions $gg\rightarrow Q\bar{Q}$ and
$q\overline{q}\rightarrow Q\bar{Q}$ are supposed to be the dominant
processes, they fail to explain important features in
hadro-production like the leading particle effects observed in $D^{\pm}$
produced in $\pi^- p$ collisions \cite{e791}, $\Lambda_c^+$ produced
in $pp$ interactions \cite{bari,isr,ruso} and others
baryons containing heavy quarks \cite{otros}, the $J/\Psi$ cross section
at large $x_F$ observed in $\pi p$ collisions \cite{na3}, etc.\\

The above mentioned effects have been explained using a two-component
model \cite{vogt} which consists of the parton fusion mechanism,
calculable in perturbative QCD, plus the coalescence of intrinsic charm
\cite{bhps-plb}.\\

Here we present an alternative mechanism, namely the conventional
recombination of valence spectator quarks with a $c$-quark present in the
sea of the initial hadron.\\ 
We show that this mechanism can explain the
production enhancement at large $x_F$ observed in charmed hadron
production in hadronic interactions.\\

We describe the $\Lambda_c^+$ production
in $pp$ interactions with a two-component model consisting of the
recombination of a diquark $ud$ with a $c$-quark from the sea of the
incident proton plus the usual parton fusion and fragmentation mechanism.
We compare the obtained results with those of the intrinsic charm
two-component model and with experimental data.\\

We also study the polarization of $\Lambda_c^+$ using the Thomas Precession
Model (TPM) \cite{dgm-prd} for both the intrinsic charm model and the
conventional recombination two component model presented here. We show
that the polarization prediction of this two models is totally different
and conclude that a measurement of the polarization can be an important
source of information to uncover the processes behind heavy 
hadron production.\\

\section{$\Lambda_c^{+}$ production in p-p collisions}

In this section we give the prediction of the two component model for
the cross section as a function of $x_F$. In section 2.1 we show the
calculation of parton fusion and fragmentation and in section 2.2
the recombination picture is presented. \\

\subsection{$\Lambda _c^{+}$ production {\it via} parton fusion}

In the parton fusion mechanism the $\Lambda _c^{+}$ is produced {\it via} the
subprocesses $q\bar{q}(gg) \rightarrow c \bar{c}$ with the subsequent
fragmentation of the $c$ quark. The inclusive $x_F$ distribution of the
$\Lambda _c^{+}$ in $pp$ collisions is given by
\cite{vbh-npb,ram-brod} 
\begin{equation} 
\frac{d\sigma^{pf}}{dx_F}=\frac{1}{2} \sqrt{s} 
\int H_{ab}(x_a,x_b,Q^2) \frac{1}{E}
\frac{D_{\Lambda_c/c} \left( z \right)}{z} dz dp_T^2 dy \: ,
\label{sig-qcd} 
\end{equation} 
where 
\begin{eqnarray}
H_{ab}(x_a,x_b,Q^2)& = & \Sigma_{a,b} \left( q_a(x_a,Q^2)
\bar{q_b}(x_b,Q^2) \right. \nonumber \\
                   &   & + \left. \bar{q_a}(x_a,Q^2) q_b(x_b,Q^2) 
\right) \frac{d \hat{\sigma}}{d
\hat{t}} \mid_{q\bar{q}} \nonumber \\
                   &   & + g_a(x_a,Q^2) g_b(x_b,Q^2) 
\frac{d \hat{\sigma}}{d \hat{t}}\mid_{gg},
\label{int-qcd}
\end{eqnarray}
with $x_a$ and $x_b$ being the parton momentum fractions, $q(x,Q^2)$ and
$g(x,Q^2)$ the quark and gluon distributions in the proton, 
$E$ the energy of the produced 
$c$-quark and $D_{\Lambda_c/c} \left( z \right)$ the 
fragmentation function measured in $e^+ e^-$ interactions. 
In eq. \ref{sig-qcd}, $p_T ^2$ is the squared transverse momentum of 
the produced $c$-quark, $y$ is the rapidity of the $\bar {c}$ quark and 
$z=x_F/x_c$ is the momentum fraction of the charm quark carried by the 
$\Lambda _{c}^{+}$. The sum in eq. \ref{int-qcd} runs over 
$a,b = u,\bar{u},d,\bar{d},s,\bar{s}$. \\

\begin{figure}[b] 
\psfig{figure=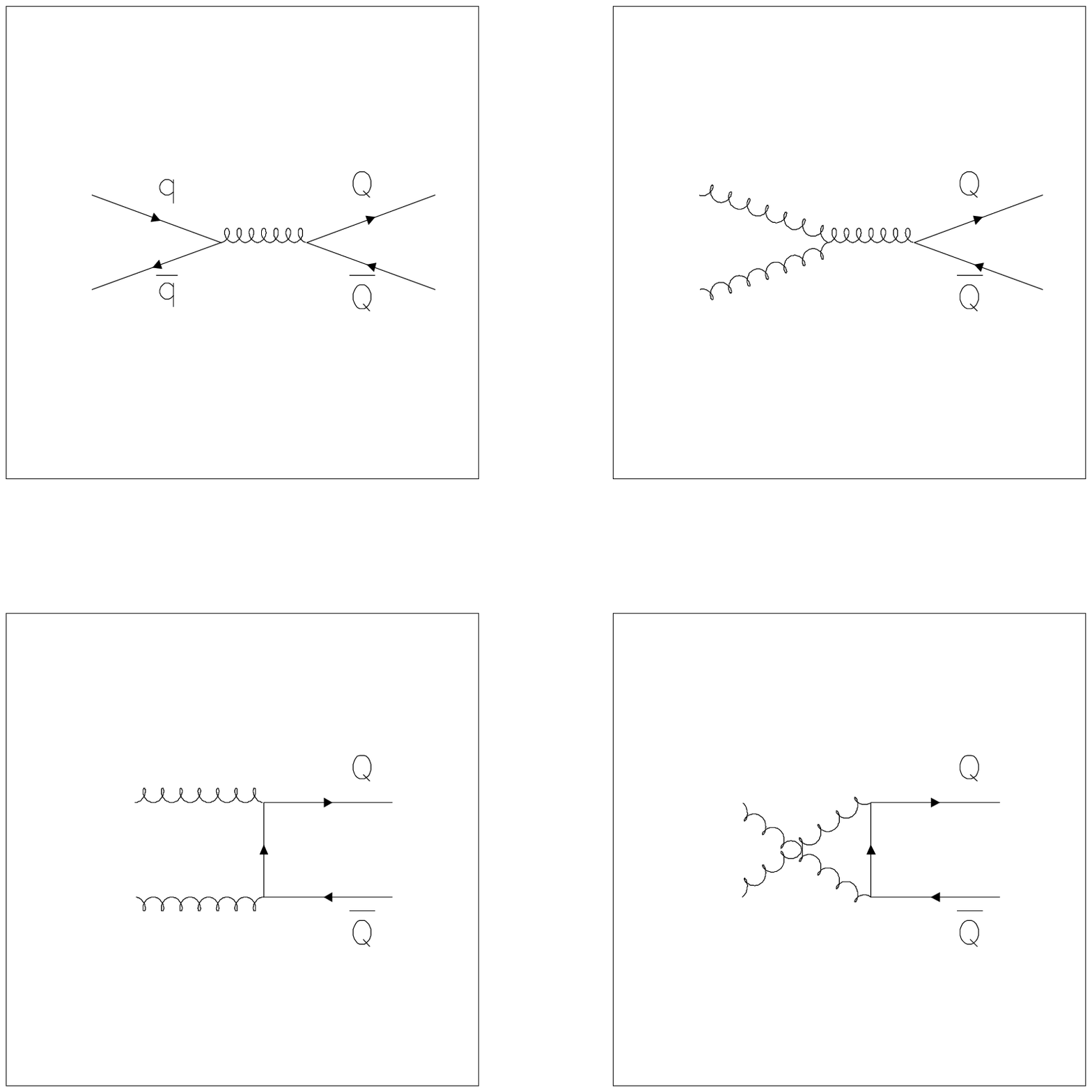,height=6.0in} 
\caption{}
\label{feynman} 
\end{figure}

We use the LO results for the elementary cross-sections $\frac{d
\hat{\sigma}}{d \hat{t}}\mid_{q \bar {q}}$ and $\frac{d \hat{\sigma}}{d
\hat{t}}\mid_{gg}$ \cite{vbh-npb}.
\begin{equation} \frac{d
\hat{\sigma}}{d \hat{t}}\mid_{q \bar {q}} = \frac{\pi \alpha _{s}^{2}
 \left( Q^2 \right)}{9 \hat{m}_{c}^{4}} \;  \frac{cosh \left( \Delta y
\right) + m_{c}^{2}/ \hat{m}_{c}^{2}} {\left[ 1+cosh \left( \Delta y
\right) \right] ^3} \label{q-antiq} 
\end{equation} 
\begin{equation}
\frac{d \hat{\sigma}}{d \hat{t}}\mid_{gg}= \frac{\pi \alpha_{s}^{2} \left(
Q^2 \right)}{96 \hat{m}_{c}^{4}} \;  \frac{8 cosh \left( \Delta y \right)
-1}{\left[ 1+cosh \left( \Delta y \right) \right]^3} \: \left[ cosh \left(
\Delta y \right)+
\frac{2m_c^2}{\hat{m}_c^2}+\frac{2m_c^4}{\hat{m}_c^4}\right], \label{gg}
\end{equation} 
where $\Delta y$ is the rapidity gap between the produced
$c$ and $\bar{c}$ quarks and $\hat{m}_c^2=m_c^2+p_T^2$.  The Feynman
diagrams involved in the calculation of eqs. \ref{q-antiq} and \ref{gg}
are shown in fig. \ref{feynman}.\\

In order to be consistent with the LO calculation of the elementary cross 
sections, we use the GRV-LO parton distribution functions \cite{gr-zpc}. 
A global factor $K \sim 2-3$ in eq. \ref{sig-qcd} takes into 
account NLO contributions \cite{ramona2}.\\

We take $m_c=1.5 \:GeV$ for the $c$-quark mass and fix the scale of the
interaction at $Q^2 = 2m_c^2$ \cite{vbh-npb}. Following \cite{vogt}, we
use two fragmentation functions to describe the
charm quark fragmentation; 
\begin{equation} 
D_{\Lambda_c/c}(z) = \delta(1-z)
\label{delta} 
\end{equation} 
and the Peterson fragmentation function
\cite{peterson} 
\begin{equation} 
D_{\Lambda_c/c}(z) = \frac{N}{z \left[
1 - 1/z - \epsilon_c/(1-z) \right]^2} \label{peter} 
\end{equation} 
with
$\epsilon_c= 0.06$ and the normalization defined by $\sum _{H} \int
D_{H/c}(z) dz = 1$. \\

\subsection{$\Lambda _c^{+}$ production {\it via} recombination}

The production of leading mesons at low $p_T$ was described by
recombination of quarks long time ago \cite{dh-plb}. The method introduced
by Das and Hwa for mesons was extended by Ranft \cite{ranft-pr} to
describe single particle distributions of leading baryons in $pp$
collisions. Recently a more sophisticated version of the
recombination model using the concept of valons \cite{hwa-prd1} has been
used to study $D^{\pm}$ asymmetries in $\pi^{-}p$ interactions
\cite{hwa-prd2}.\\

In recombination models it is assumed that the outgoing leading hadron is 
produced in the beam fragmentation region through the recombination of 
the maximum number of valence- and the minimun number of sea- quarks.
Thus $\Lambda _c^+$'s produced in $pp$ collisions are formed
by an $u$ and a $d$ valence quarks and a $c$ sea quark from the incident
proton.
Contributions involving the recombination of more than
one sea flavor are neglected.\\ The invariant inclusive $x_F$ distribution
for leading baryons is given by 
\begin{equation} 
\frac{2 E}{\sigma \sqrt{s}} 
\frac{d\sigma^{rec}}{dx_F}=
\int_0^{x_F}\frac{dx_1}{x_1}\frac{dx_2}{x_2}
\frac{dx_3}{x_3}F_3\left( x_1,x_2,x_3\right) 
R_3\left( x_1,x_2,x_3,x_F\right),
\label{rec-cs} 
\end{equation} 
where $x_i$, $i=1,2,3$, is the momentum
fraction of the $i^{th}$ quark, $F_3 \left( x_1,x_2,x_3 \right) $ is the
three-quark distribution function in the incident hadron and $R_3\left(
x_1,x_2,\right.$ $\left.x_3,x_F\right) $ is the three-quark recombination
function.\\
The $\Lambda_c^+$ production by recombination is shown in schematic form
in fig. 2.\\

We use a parametrization containing explicitely the single quark
distributions for the three-quark distribution function 
\begin{equation}
F_3 \left( x_1,x_2,x_3 \right) = \beta
F_{u,val}\left(x_1\right)F_{d,val}\left(x_2\right)F_{c,sea}\left(x_3\right)
\left(1-x_1-x_2-x_3\right)^{\gamma} \label{3-quark} 
\end{equation} 
with
$F_{q}\left(x_i\right) = x_iq\left(x_i\right)$ and $F_u$ normalized to one
valence $u$ quark. The parameters $\beta$ and $\gamma$ are constants fixed
by the consistency condition 
\begin{eqnarray} 
F_{q}\left(x_i\right) & = & \int_0^{1-x_i}dx_j \int_0^{1-x_i-x_j}dx_k 
\:F_3 \left( x_1,x_2,x_3 \right),\nonumber \\
                      &   &i,j,k = 1,2,3 
\label{consist} 
\end{eqnarray} 
for the valence quarks of the incoming protons as in ref.
\cite{ranft-pr}.\\

We use the GRV-LO parametrization for the single quark distributions in
eqs. \ref{3-quark} and \ref{consist}. It must be noted that since the
GRV-LO distributions are functions of $x$ and $Q^2$, our $F_3 \left(
x_1,x_2,x_3 \right)$ also depends on $Q ^2$. \\

In contrast with the parton fusion calculation, in which the scale $Q^2$
of the interaction is fixed at the vertices of the appropriated Feynman
diagrams, in recombination there is not clear way to fix the value of the
parameter $Q^2$, which in this case is not properly a scale parameter and 
should be used to give the content of the recombining 
quarks in the initial hadron.\\

Since the charm content in the proton sea increases rapidly for $Q^2$
growing from $m_c^2$ to $Q^2$ of the order of some $m_c^2$'s when it
becomes approximately constant, we take $Q^2 = 4 m_c^2$, a conservative
value, but sufficiently far from the charm threshold in order to avoid a
highly depressed charm sea which surely does not represent the real
charm content of the proton. At this value of $Q^2$ we found that the
condition of eq. \ref{consist} is fulfilled approximately with 
$\gamma = -0.1$ and $\beta = 75$.  We have verified that the 
recombination cross section does not change appreciably at higher values 
of $Q^2$.\\

The value of $Q^2$ in recombination could be different from the scale of the 
interaction in parton fusion. In fact, in the later the scale of the 
interaction should be chosen to be of
the order of the hard momentum scale while in recombination, the value of
$Q^2$ must be chosen in such a way that the true content of quarks be
present to form the outgoing hadron, allowing in this way a more higher
value of this parameter.\\

For the three-quark recombination function for $\Lambda_c^+$ production we
take the simple form \cite{ranft-pr} 
\begin{equation} 
R_3\left(x_u,x_d,x_c\right) =\alpha \frac{x_ux_dx_c}{x_F^2}\delta \left(
x_u+x_d+x_c-x_F\right) \: , \label{eq7} 
\end{equation} 
with $\alpha$ fixed by the condition 
$\int_0^1 dx_F (1/\sigma)d\sigma^{rec}/dx_F = 1$,
where $\sigma$ is the cross section for $\Lambda_c^+$'s inclusively
produced in $pp$ collisions. Of course, with this choice for the parameter
$\alpha$ we will not be able to give a prediction for the total cross
section for $\Lambda_c^+$ production.
The recombination model of eq. \ref{rec-cs} can only give predictions
on cross sections relative to the cross section for a known process. 
Thus, we could take a different point of view and chose $\alpha$ in order
to fit the inclusive $x_F$ distribution for a given reaction, {\it e.g.}
$pp \rightarrow p+X$, and expect that the relative normalizations of the
single quark distributions account for the desired proportion of another
processes cross sections relative to the $pp \rightarrow p+X$ cross section,
as was made in ref. \cite{ranft-pr}. Instead of it, and since in the case
of $pp \rightarrow \Lambda_c^+ + X$ we don't know the total cross section
accurately, we simply fix $\alpha$ in order to normalize to $1$ the
$\Lambda_c^+$'s $x_F$ distribution as given by eq. \ref{rec-cs}, lossing
the possibility to give predictions on the total inclusive $\Lambda_c^+$
cross section in recombination.  Other forms than eq. \ref{eq7} for the
recombination function have been considered in the literature
\cite{hwa-prd1,hwa-prd2,recombinacion}. We have observed
that the shape of the inclusive cross section of eq. \ref{rec-cs} is
practically insensitive to the form of the recombination function as was 
pointed out in conection with $\Lambda_0$ production in $pp$ collisions 
\cite{ranft-pr}.\\

Using eqs. \ref{3-quark} and \ref{eq7} the invariant $x_F$ distribution
for $\Lambda_c^+$'s produced inclusively in $pp$ collisions can be written
as 
\begin{eqnarray} 
\frac {2 E}{\sqrt {s}\sigma}
\frac{d\sigma_{\Lambda_c^+}^{rec}}{dx_F} & = & \alpha 75
\frac{(1-x_F)^{-0.1}}{x_F^2} \int_0^{x_F} dx_1 F_{u,val}(x_1) \nonumber \\
& & \times \int_0^{x_F-x_1} dx_2 F_{d,val}(x_2) F_{c,sea}(x_F-x_1-x_2) \:
, \label{rec-lambda} 
\end{eqnarray} 
where we integrate out over $x_3$.
The parameter $\sigma$ is fixed with experimental data.\\

The inclusive production cross section of the $\Lambda_c^{+}$ is obtained
by adding the contribution of recombination (eq. \ref{rec-lambda}) to the
QCD processes of eq. \ref{sig-qcd},
\begin{equation}
\frac{d\sigma^{tot}}{dx_F} = \frac{d\sigma^{pf} }{dx_F} +
\frac{d\sigma^{rec}}{dx_F}.  \label{sig-tot2} 
\end{equation}

The resulting inclusive $\Lambda_c^{+}$ production cross section
$d\sigma^{tot}/dx_F$ is plotted in fig. \ref{fig-sig} a) and b) using the two
fragmentation function of eqs. \ref{delta} and \ref{peter} and compared
with experimental data in $pp$ collisions from the ISR \cite{isr}. As we
can see, the shape of the experimental data is very well described by our
model. We use a factor $\sigma = 0.92 (0.72) \mu b$ for
Peterson (delta) fragmentation respectively. 

Although we are not able to explain the normalization of the ISR data (we
need a global factor of 1000 in eq. \ref{sig-tot2}), its $x_F$ dependence
is well described by our model.\\

\section{$\Lambda _c^{+}$ production by IC coalescence and fragmentation}

In a similar approach R. Vogt {\it et al.} \cite{vogt} calculated the
$\Lambda _c^{+}$ production in $pp$ and $\pi p$ collisions.  The two
component model they use consists of a parton fusion mechanism plus
coalescence of the intrinsic charm in the proton.\\

In ref. \cite{vogt} it is assumed that $\Lambda _c^{+}$ 's can be produced
by coalescence of a $c$ quark with a $ud$ diquark when the coherence of a
$|uudc \bar{c} \rangle$ Fock state of the proton breaks due to the
inelastic interaction with the target in $pp \rightarrow \Lambda_{c}^{+} +
X$ reactions. For $\Lambda_{c}^{+}$ production in $\pi^{-} p$ collisions,
a $| \bar{u}dc \bar{c} \rangle$ Fock state of the pion is considered.\\

The frame independent probability distribution of a five particle
state of the proton is \cite{vogt,bhps-plb} 
\begin{equation}
\frac{dP^{IC}}{dx_udx_{u'}...dx_{\bar c}} = N_5 \alpha_s^4\left(M_{c
\bar{c}}^2\right) \frac{\delta \left(1-\Sigma _{i=u}^{\bar c} x_i
\right)}{\left( m_p^2 - \Sigma _{i=u}^{\bar c} \hat {m}_i^2 x_i \right)^2}
\: , 
\label{ic-prob} 
\end{equation} 
with $N_5$ normalizing the $\left|uudc \bar{c} \right\rangle$ state 
probability.\\

The intrinsic charm $x_F$ distribution is related to $P^{IC}$ by
\cite{vogt}
\begin{eqnarray} 
\frac{d\sigma^{IC} }{dx_F} & = & \sigma^{in}_{pp} 
\frac{\mu ^2}{4 \hat{m}_c^2} 
\int_0^1dx_udx_{u^{\prime}}dx_ddx_cdx_{\bar{c} }\delta 
\left( x_F-x_u-x_d-x_c\right) \nonumber \\
                           &   & \times 
\frac{dP^{IC}}{dx_udx_{u'}...dx_{\bar c}} 
\label{ic-cs} 
\end{eqnarray} 
where the factor $\mu^2 / 4 \hat{m}_c^2$ is the result of the
soft interaction needed to break the coherence of the Fock state.
$\sigma^{in}_{pp}$ is the inelastic $pp$ cross section.\\

In ref \cite{vogt} an attempt is made to fix the soft scale parameter at
$\mu ^2 \sim 0.2 \: GeV^2$ by the assumption that the diffractive fraction
of the total cross section is the same for charmonium and charmed hadrons.
In this way they obtain $\sigma^{IC}_{pN} \simeq 0.7 \: \mu b$ in $pN$
interactions at $200 \: GeV$ with $P^{IC} = 0.3 \%$.\\

Since we want to compare the prediction of the IC model for $\Lambda _c^+$
production in $pp$ collisions at the ISR energy $\sqrt {s} = 63\: GeV$,
we prefer to fix the unknown product $\sigma^{in}_{pp} \mu ^2$ with the
experimental data.\\

The total inclusive cross section for $\Lambda_c^+$ production in $pp$
collisions is then given by, 
\begin{equation} 
\frac{d\sigma^{tot}}{dx_F} = \frac{d\sigma^{pf} }{dx_F} 
+r \frac{d\sigma^{IC}_{rec}}{dx_F} + \frac{d\sigma^{IC}_{frag}}{dx_F} 
\label{ic-tcs} 
\end{equation} 
where $r$ represents the fraction of IC production 
with respect to parton fusion (see ref. \cite{vogt}). 

The prediction of the IC two-component model is plotted in fig.
\ref{fig-sig} c) and d) for Peterson and delta fragmentation
respectively and compared to the prediction of the recombination
two-component model and experimental data from the ISR \cite{isr}. We use
$\hat {m}_c = 1.8 \:GeV$ as quoted in ref. \cite{vogt} and
$r\sigma^{in}_{pp}\mu^2 = 270 (320)$ for delta
(Peterson) fragmentation in order to fit the experimental data with the IC
two-component model. It must be mentioned that, as with the recombination
two-component model, the IC two component-model can not explain the
abnormally high normalization of the ISR data and a global factor of 1000
is needed in eq. \ref{ic-tcs}.\\

\section{The Thomas Precession Model and the $\Lambda_c^{+}$
polarization}

In the TPM the polarization of hadrons is a consequence of the Thomas
precession during the recombination process \cite{dgm-prd}.\\
 
A $\Lambda $ baryon is formed by the recombination of a $ud$ diquark in a
spin state $j=m=0$ and a $Q$ quark originally present in the sea of the
projectile, so the spin of the $\Lambda$ is carried entirely by the $Q$
quark.\\

Since the recombining quarks in the projectile must carry a fraction of
the outgoing hadron's transverse momentum, the Thomas precession
appears due to the change in the longitudinal momentum of quarks as they
pass from the projectile to the final hadron.\\

In particular, if the $\Lambda_{c}^{+}$ is produced through Valence
Valence Sea (VVS) recombination, since the $x$-distribution of sea quarks is
very step, the $\Lambda_{c}^{+}$ must get most of its momentum from the
valence $ud$ diquark. In this case the $c$ sea-quark must be accelerated
in passing from the proton sea to the produced $\Lambda_{c}^{+}$ and the
later must be negatively polarized due to the Thomas precession.\\

On the other hand, if the $\Lambda_{c}^{+}$ is produced by the coalescence
mechanism, since the $x$-distribution of the intrinsic charm is hard, the
$\Lambda_{c}^{+}$ gets most of its momentum from the intrinsic $c$-quark,
which must be decelerated in the recombination process and consequently
the $\Lambda_{c}^{+}$ is produced with positive polarization
\cite{mh-plb}.\\

The amplitude for the production of a $\Lambda$ of spin $\vec{s}$ is
proportional to $(\Delta E + \vec{s} \cdot \vec{\omega_T})$, where 
$\Delta E$ represents the change in energy in going from the quarks to 
the final state in absence of spin effects and 
$\vec{\omega_T} \sim \gamma/(1+\gamma) \vec{F} \times \vec{\beta}$ is the 
Thomas frequency.\\

The polarization in the TPM si given by 
\begin{equation} 
P\left(p\rightarrow \Lambda \right) =
-\frac{12}{\Delta xM^2}\frac{\left[ 1-3\xi
\left( x_F\right) \right] }
{\left[ 1+3\xi \left( x_F\right) \right] ^2}p_{T\Lambda },
\label{polar} 
\end{equation} 
where 
\begin{equation}
M^2=\left[ \frac{m_D^2+p_{TD}^2}{1-\xi \left( x_F\right) }+
\frac{m_q^2+p_{Tq}^2}{\xi \left( x_F\right) }-
m_\Lambda ^2-p_{T\Lambda}^2\right], 
\label{eq2} 
\end{equation} 
where $m_D$ and $m_q$ are the masses 
and $p_{TD}$ and $p_{Tq}$ the transverse 
momentums of the $ud$ diquark and $c$ quark respectively, $m_\Lambda $ and 
$p_{T\Lambda }$ the mass and transverse momentum of the $\Lambda_c$, and 
$\xi \left( x_F \right) = \langle x_Q \rangle / x_F$ 
the average momentum fraction of the $Q$ quark in the projectile.\\

To give a quantitative prediction for the polarization, DeGrand and Miettinen
take a parametrization for $\xi \left( x_F \right)$ and obtain a good 
description of experimental data \cite{dgm-prd}.
We explicitly compute $\xi \left( x_F \right)$ \cite{mh-plb,hmms-plb} 
using a recombination model 
\begin{equation}
\langle \xi \rangle \left( x_F\right) =
\frac{\left\langle x_q\right\rangle }{x_F}=
\frac{\int dx_qx_q\frac{d\sigma }{dx_qdx_F}}{x_F \frac{d\sigma }{dx_F}}  
\label{eq3}
\end{equation}
where $x_F$ is the Feynman $x$ defined by $x_{F} = 2 p_{L} / \sqrt s$.\\
The production of $\overline{\Lambda} _{c}^{-}$ would have a recombination
component which contributes with a smaller fraction to the total cross
section. This is a consequence of the fact that in a proton beam the
production of the $\overline{\Lambda} _{c}^{-}$ takes place through the
recombination of sea quarks only.\\ 

In the production mechanism via intrinsic charm 
the $\overline{\Lambda} _{c}^{-}$ can only be 
produced by the $c$-quark fragmentation in a five 
particle Fock state. If a nine particle Fock state is considered 
$\left| uudc\overline{c}u\overline{u}d\overline{d} \right\rangle$
as pointed out in \cite{ram-brod}, the $x_c$ distribution would not be 
as hard as in the five particle Fock state. Therefore the recombination
would not decelerate the $\bar{c}$ quark 
as it happens with $c$ quark in the $\Lambda_c$ formation via 
recombination.\\
Henceforth, the two mechanisms predict zero polarization for the 
$\overline{\Lambda} _{c}^{-}$.

\subsection{The polarization in a two component model}
%
The polarization is defined as 
\begin{equation} 
P\left( x_F\right) = \left[\frac {d\sigma_\uparrow}{dx_F}-
\frac{d\sigma_\downarrow}{dx_F}\right] / \frac{d\sigma}{dx_F},
\label{pol-gen}
\end{equation} 
where $d\sigma_\uparrow/dx_F$,
$d\sigma_\downarrow/dx_F$ and $d\sigma/dx_F$ are the spin up, spin
down and total cross section respectively.
In a scenario where $\Lambda_c^+$s originate in two different processes
one must take into account the different contributions to the polarization.\\

The polarization of baryons in the TPM  is the result of 
the recombination process.
$\Lambda_c^+$ produced by parton fusion and fragmentation has been shown to 
have a very small polarization \cite{mh-plb} which will be neglected here.
We have,
\begin{equation} 
\frac{d\sigma^{tot}}{dx_F} =
\frac{d\sigma_{\downarrow}^{rec}}{dx_F} +
\frac{d\sigma_{\uparrow}^{rec}}{dx_F} + \frac{d\sigma^{pf}}{dx_F} \: .
\label{ntot} 
\end{equation} 
We now define the fraction of $\Lambda_c^+$ produced by parton fusion
of the total of $\Lambda_c^+$s produced as
\begin{equation}
g\left(x_F\right) =
\left[\frac{d\sigma^{pf}}{dx_F}\right]/\left[\frac{d\sigma^{tot}}{dx_F}\right]
\label{rxf} 
\end{equation} 
hence, from eq. \ref{sig-tot2} 
\begin{equation} 
\frac{d\sigma ^{pf}}{dx_F} =
\frac{g\left(x_F\right)}{1-g\left(x_F\right)} \frac{d\sigma ^{rec}}{dx_F}
\: . \label{sig-pf} 
\end{equation} 
Using eq. \ref{sig-pf} we can rewrite eq. \ref{ntot} as 
\begin{equation} 
\frac{d\sigma}{dx_F} =
\frac{1}{1-g\left(x_F\right)}
\left(\frac{d\sigma_{\downarrow}^{rec}}{dx_F} +
\frac{d\sigma_{\uparrow}^{rec}}{dx_F} \right) 
\label{sig-tot}
\end{equation} 
and with the help of eqs.  \ref{ntot} and
\ref{sig-tot} we obtain the polarization 
\begin{equation}
P\left(x_F\right) =
\left[1-g\left(x_F\right)\right]\left[\frac{d\sigma_{\uparrow}^{rec}}{dx_F}
-\frac{d\sigma_{\downarrow}}{dx_F}^{rec}\right]/\left[
\frac{d\sigma_{\uparrow}^{rec}}{dx_F}
+ \frac{d\sigma_{\downarrow}^{rec}}{dx_F}\right]
\label{pol1}
\end{equation} 
or using eq. \ref{polar} 
\begin{equation}
P\left(x_F\right) = 
- \left[1-g\left(x_F\right)\right]\frac{12}{\Delta xM^2}
\frac{\left( 1-3\xi \left( x_F\right) \right) }{\left( 1+3\xi \left(
x_F\right) \right) ^2 }p_{T\Lambda } 
\label{pol2} 
\end{equation} 
where $M^2$ is given by eq. \ref{eq2}. A similar reasoning gives the same eq.
\ref{pol2} for the polarization of $\Lambda_c^{+}$ produced by intrinsic
charm coalescence. \\
The $g\left(x_F\right)$ for the two models are shown in fig. \ref{frac}.

%
\subsection{The $\Lambda _c^{+}$ polarization}

In the coalescence of the intrinsic charm of the proton with 
quarks in a Fock state the $x_F$ distribution of the $\Lambda_c^+$ is 
given by eq. \ref{ic-cs}. So that we can write
\begin{equation}
\frac{d\sigma }{dx_cdx_F}= \left( x_{F}-x_c \right)\int_0^{1-x_F}dx_u
\left( \frac{x_{c} \left( 1-x_u-x_F\right) }{x_c+1-x_u-x_F}\right) ^2.
\label{xc-ic}
\end{equation}

If the recombination mechanism occurs among the diquark $ud$ in the 
proton and the $c$ quark from the proton's sea then, from eq. \ref{rec-lambda},
we have
\begin{equation}
\frac{d\sigma }{dx_cdx_F}=
\int_0^{x_F-x_c}dx_u\int_0^{x_F-x_c-x_u}dx_d x_c F_3\left(
x_u,x_d,x_c\right) R_3\left( x_u,x_d,x_c\right) \: .  
\label{xc-rec}
\end{equation}
In order to calculate the polarization of the produced $\Lambda_c^+$'s, 
we need the mean value of $\xi^{IC\left(rec\right)}\left(x_F\right)$ 
which is obtained replacing eqs. \ref{xc-ic} and \ref{xc-rec} into 
eq. \ref{eq3} for each case respectively. 
The $\xi(x_F)$ for these two processes are shown in fig. \ref{xi-ic-rec}.\\

We calculate the $\Lambda _c^{+}$ polarization given by eq. \ref{pol2} for
each production mechanism separately using $\xi \left( x_F\right) $ as
given by eq. \ref{eq3} with eqs. \ref{xc-rec} and \ref{xc-ic} for
recombination and coalescence respectively.\\

The $\Lambda_c^+$ polarization is plotted in fig. \ref{fig-pol} for the
two models; parton fusion plus recombination and parton fusion plus
IC-coalescence. As in ref. \cite{dgm-prd} we use $ m_D=\frac 23 \:GeV$ , 
$\left\langle p_T^2\right\rangle_{D,c}=\frac 14p_{T\Lambda }^2+
\left\langle k_T^2\right\rangle $ with
$\left\langle k_T^2\right\rangle =0.25 \: GeV^2$ and $\Delta x=5 \:
GeV^{-1}$ . We take $m_c=1.5 \: GeV $ and $m_{\Lambda _c^{+}}=2.285 \:
GeV$ . 

\section{Conclusions}

We  studied the $\Lambda_c^+$ production in $pp$ collisions in the
framework of two component models. We used two different versions of
the two component model and compared them. As we can see in fig.
\ref{fig-sig} c) and d), both versions can describe the shape of the $x_F$
distribution for $\Lambda_c^+$'s produced in $pp$ collisions, but none of
them can describe the abnormally high normalization of the ISR data quoted
in ref. \cite{isr}. At this point it is useful to mention that the
normalization of the ISR data has been object of study in other
publications (see {\it e.g.} ref. \cite{rusos2}) and none of them was able
to explain it. This discrepancy between theory and experiment does not
exist for charmed meson production, which is well described both in shape
and normalization with the parton fusion mechanism plus intrinsic charm
coalescence \cite{ram-brod}. An analysis of charmed meson production using
parton fusion and recombination is forthcoming \cite{nos}.\\

In $pp$ interactions
the intrinsic charm component seems to be consistent with $1\%$ or
less of the total charm sea as stressed in ref. \cite{ramona}.
This suggests that the recombination of valence and $c$-sea quarks may 
be an alternative mechanism to explain the enhancement in the $x_F$
distribution at large values.\\

The Thomas Precession Model predicts very different
results for the $\Lambda_c^+$ polarization in $pp$ collisions
for the production mechanisms discussed. Unfortunately, the only known
experimental measurement of the $\Lambda_c^+$ polarization in $pp$ collisions
\cite{ruso} is not able to give the sign, then we can not obtain
definitive conclusions about the production mechanisms of the
$\Lambda_c^+$. However, it is interesting to note that the measurement
quoted in ref. \cite{ruso} seems to be compatible with a high value for
the $\Lambda_c^+$ polarization, despite the large experimental errors,
which may indicate that the TPM is not adequated to describe the
polarization of hadrons containing heavy quarks. At this respect, it is
interesting to note that the TPM predicts small values for the
$\Lambda_c^+$ polarization because of its big mass,
(see eqs. \ref{polar} and \ref{eq2}). Moreover, the fact that only a 
fraction of the total number of $\Lambda_c^+$s produced originated
in a recombination process, supresses the polarization even more. \\

Another sensible test that can give information about the production
mechanisms of hadrons containing heavy quarks is the measurement of hadron
anti-hadron production asymmetries. In fact, since the parton fusion
mechanism predicts a very small asymmetry at NLO, the 
asymmetry observed experimentally in charmed
hadron production must be produced by conventional recombination or the 
intrinsic charm coalescence mechanism.\\

The $\Lambda_c/\bar{\Lambda_c}$ asymmetry has been calculated in $pp$
collisions using the two-component model with intrinsic charm coalescence
\cite{vogt}. In \cite{vogt}, a factor $r=100$ is needed to 
describe properly the asymmetry compared to PHYTIA.  Such very high value 
for the parameter $r$ seems unnatural because it produces an enlargement 
of the intrinsic charm cross section which can not be easily explained.
In addition, the intrinsic charm two-component model is not able to
describe the shape of the experimentally observed asymmetry in 
$D^{\pm}$ production in $\pi^- p$ interactions \cite{e791}.

\section*{Acknowledgements} 
J.M and G.H. wish to thank to Centro Brasileiro
de Pesquisas F\'{\i}sicas (CBPF), for the hospitality extended to them. We
thank H.R. Christiansen for enlightening discussions during the completion
of this work.

\newpage 
\section*{Figure Captions} 
\begin{itemize} 
\item [Fig. 1:]
Feynman diagrams involved in the calculation of the LO elementary cross
sections of hard processes in parton fusion.
\item [Fig. 2:]
Recombination picture: a $c$-quark joints a $ud$-diquark to 
build a $\Lambda_c^+$. 
\item [Fig. 3:]
\begin{itemize} 
\item [a)] $x_F$ distribution predicted by parton fusion
with a Peterson fragmentation function (dashed line) and the sum
of the QCD and the conventional recombination contributions (solid line). 
\item [b)] $x_F$ distribution predicted by parton fusion
with a delta fragmentation function (dashed line) and the sum
of the QCD and the conventional recombination contributions (solid line).
\item [c)] $x_F$ distribution
predicted by parton fusion plus recombination (full line) and parton
fusion plus IC coalescence (dashed line) with the Peterson fragmentation
function. 
\item [d)] 
Same as in (c) for delta function recombination.\\ 

\noindent
Experimental data (black dots) are taken from ref. 3.
\end{itemize}

\item [Fig. 4:] Fraction $g(x_F)$ of $\Lambda_c^+$ produced by 
parton fusion and conventional recombination (solid line) and IC recombination
(dashed line).
In (a) the delta and (b) the Peterson fragmentation function was
used to describe the hadronization.
\item [Fig. 5:] 
$\xi (x_F)$ obtained from the conventional recombination model (solid line) 
and from IC coalescence (dashed line). 
\item [Fig. 6:] 
$\Lambda_c^+$ polarization as a function of $x_F$ at $p_T = 0.5 \:
GeV$ for parton fusion plus recombination (lower curves) and parton fusion
plus IC coalescence (upper curves):\\
We show both when a Peterson (solid line)  
and a delta (dashed line) fragmentation function was used. 
In a similar way for the upper curves a Peterson (dotted line)
and a delta (dot-dashed line) fragmentation function was used in the
hadronization. 
\end{itemize} 
\newpage 
\begin{figure}
\begin{picture}(100,250)(-90,-150)
\put(-10,-10){\framebox(220,140)}
\multiput(50,40)(0,0.1){600}{\circle*{20}}
\put(0,110){\vector(2,-1){27}}
\put(0,110){\line(2,-1){55}}
\put(23,110){{\bf p}}
\put(0,30){\vector(2,1){27}}
\put(0,30){\line(2,1){55}}
\put(23,30){{\bf p}}
\put(55,95){\vector(1,0){42}}
\put(55,95){\line(1,0){80}}
\put(55,90){\vector(1,0){42}}
\put(55,90){\line(1,0){80}}
\put(85,100){{\bf ud}}
\put(55,60){\vector(4,1){42}}
\put(55,60){\line(4,1){80}}
\put(95,61){{\bf c}}
\multiput(45,60)(0,-4){5}{\vector(2,-1){40}}
\multiput(45,60)(0,-4){5}{\line(2,-1){80}}
\put(140,5){{\bf X}}
\multiput(140,70)(0,0.1){300}{\circle*{20}}
\put(145,85){\vector(1,-0){27}}
\put(145,85){\line(1,-0){55}}
\put(165,90){$\Lambda_c^+$}
\end{picture}
\label{figrec}
\caption{}
\end{figure}
\begin{figure}[b]
\psfig{figure=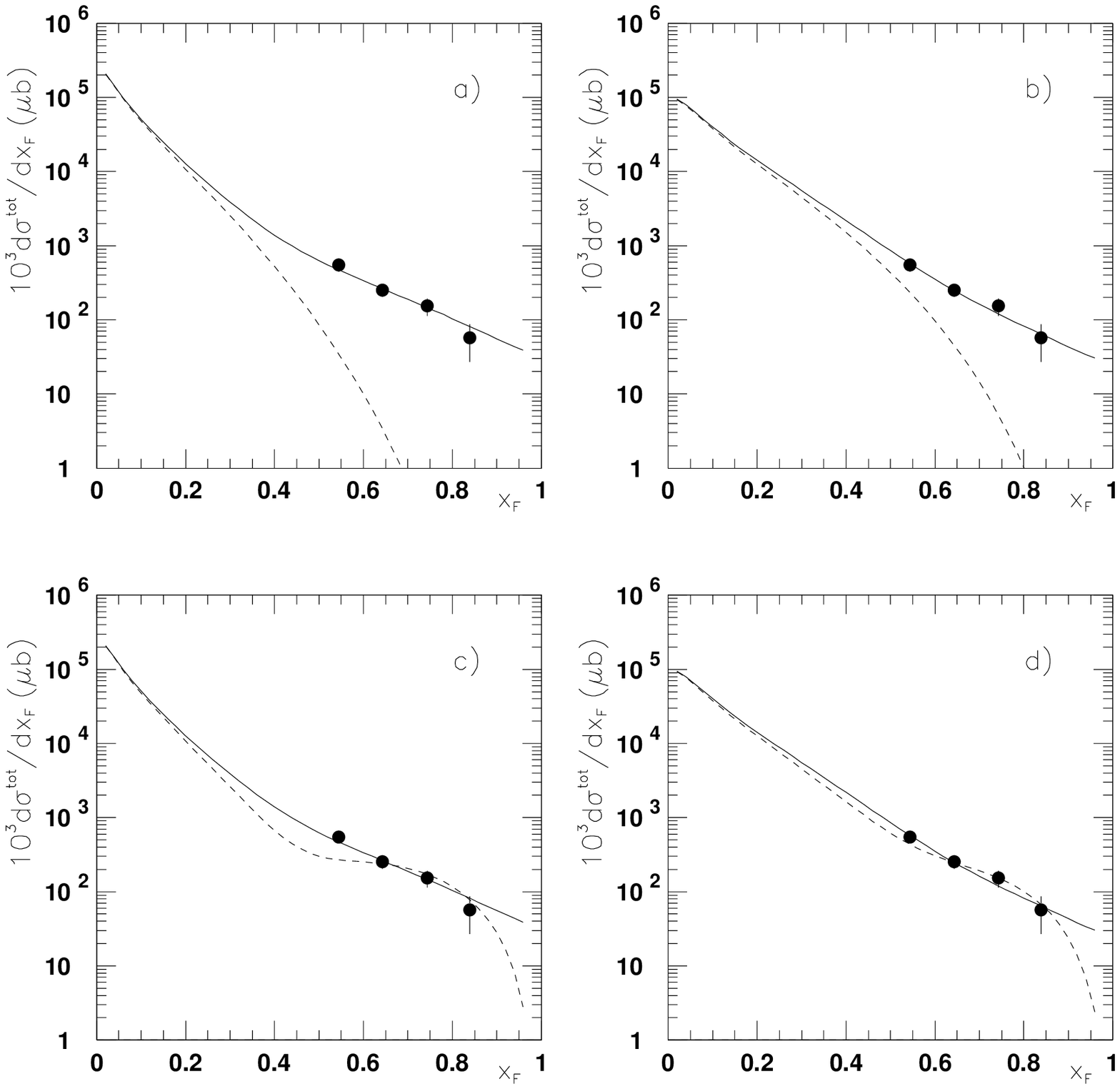,height=6.0in} 
\caption{} 
\label{fig-sig}
\end{figure} 
\begin{figure}[b]
\psfig{figure=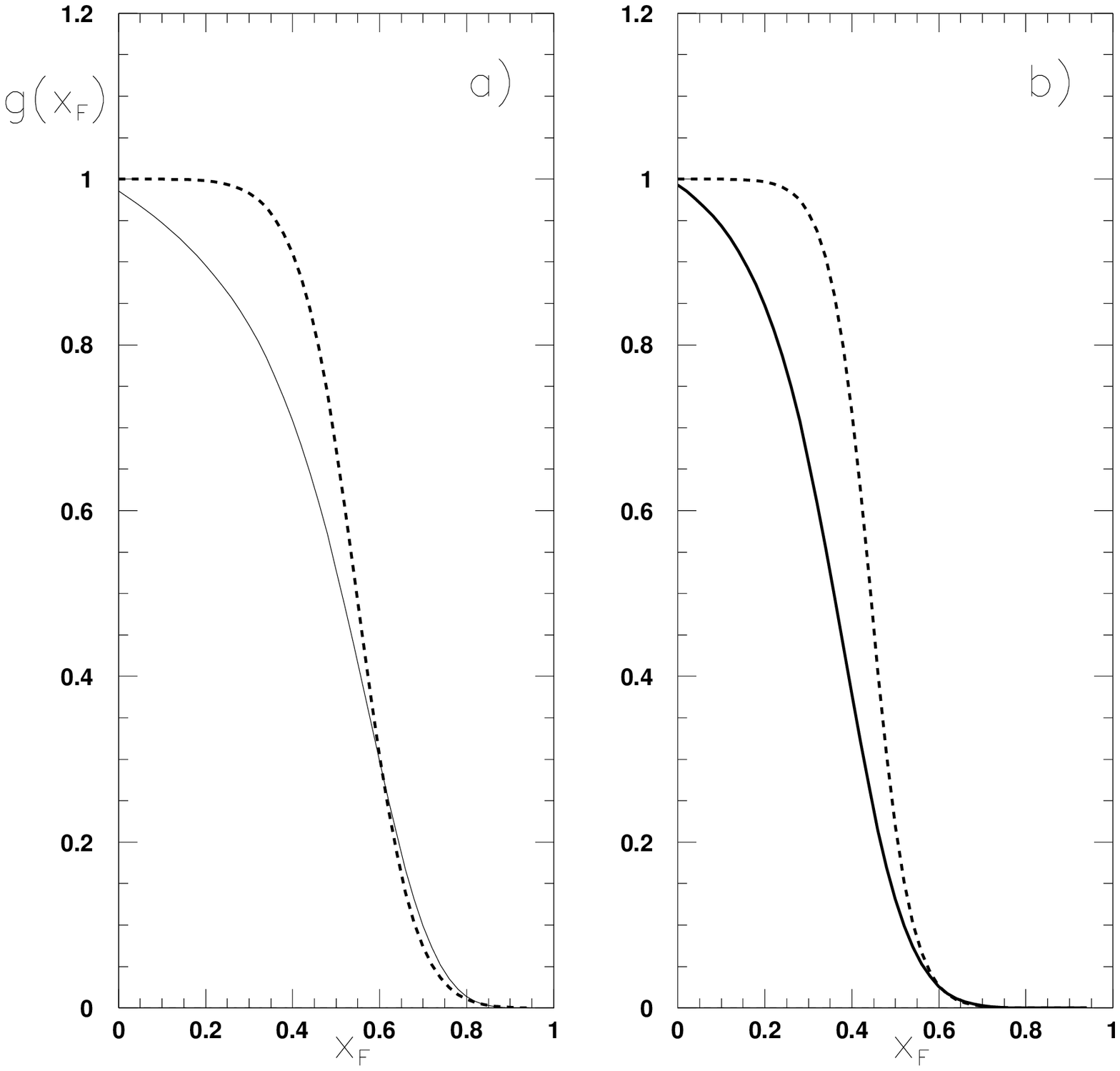,height=6.0in}
\caption{}
\label{frac}
\end{figure} 
\begin{figure}[b] 
\psfig{figure=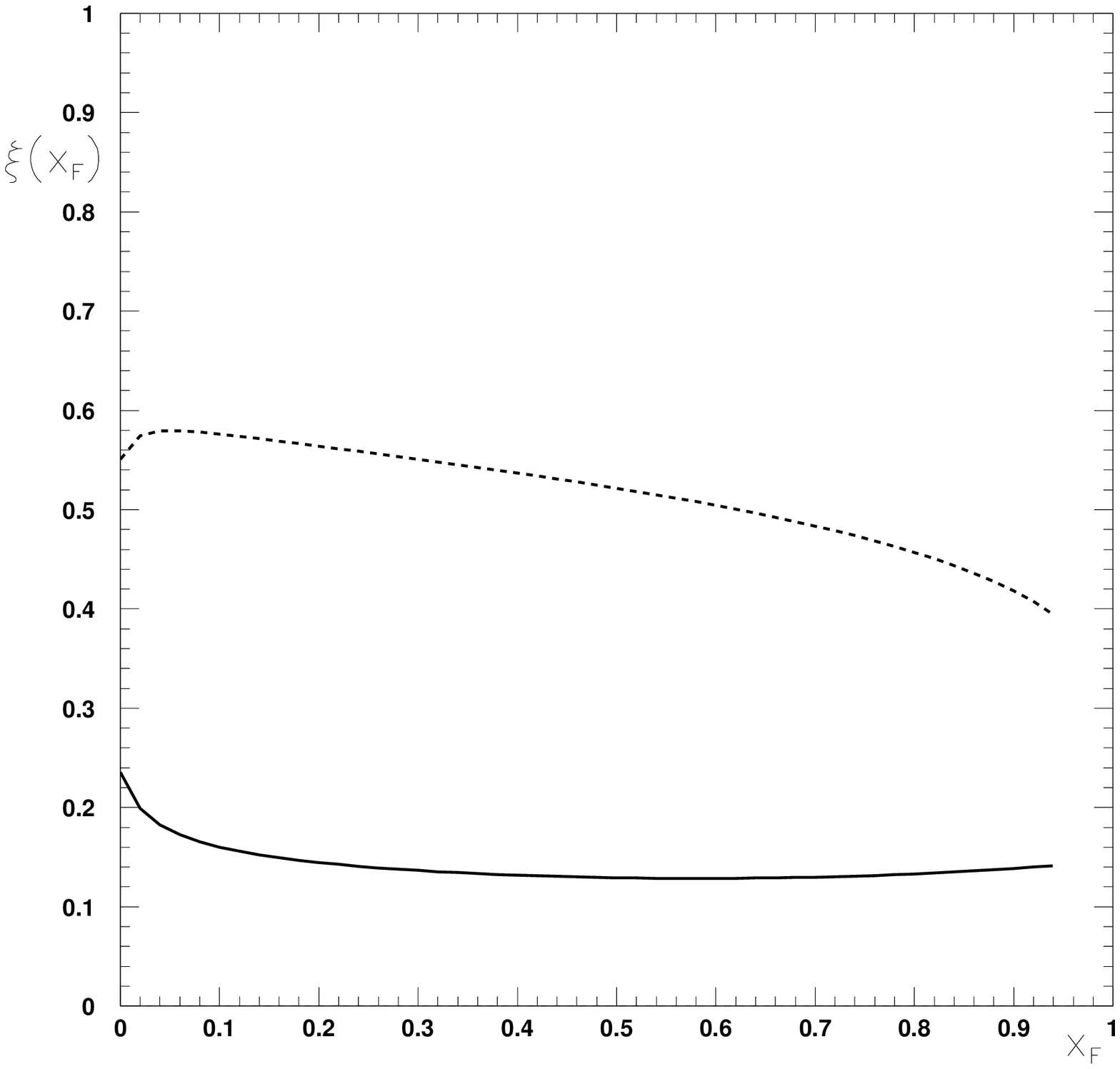,height=6.0in}
\caption{} 
\label{xi-ic-rec} 
\end{figure} 
\begin{figure}[b]
\psfig{figure=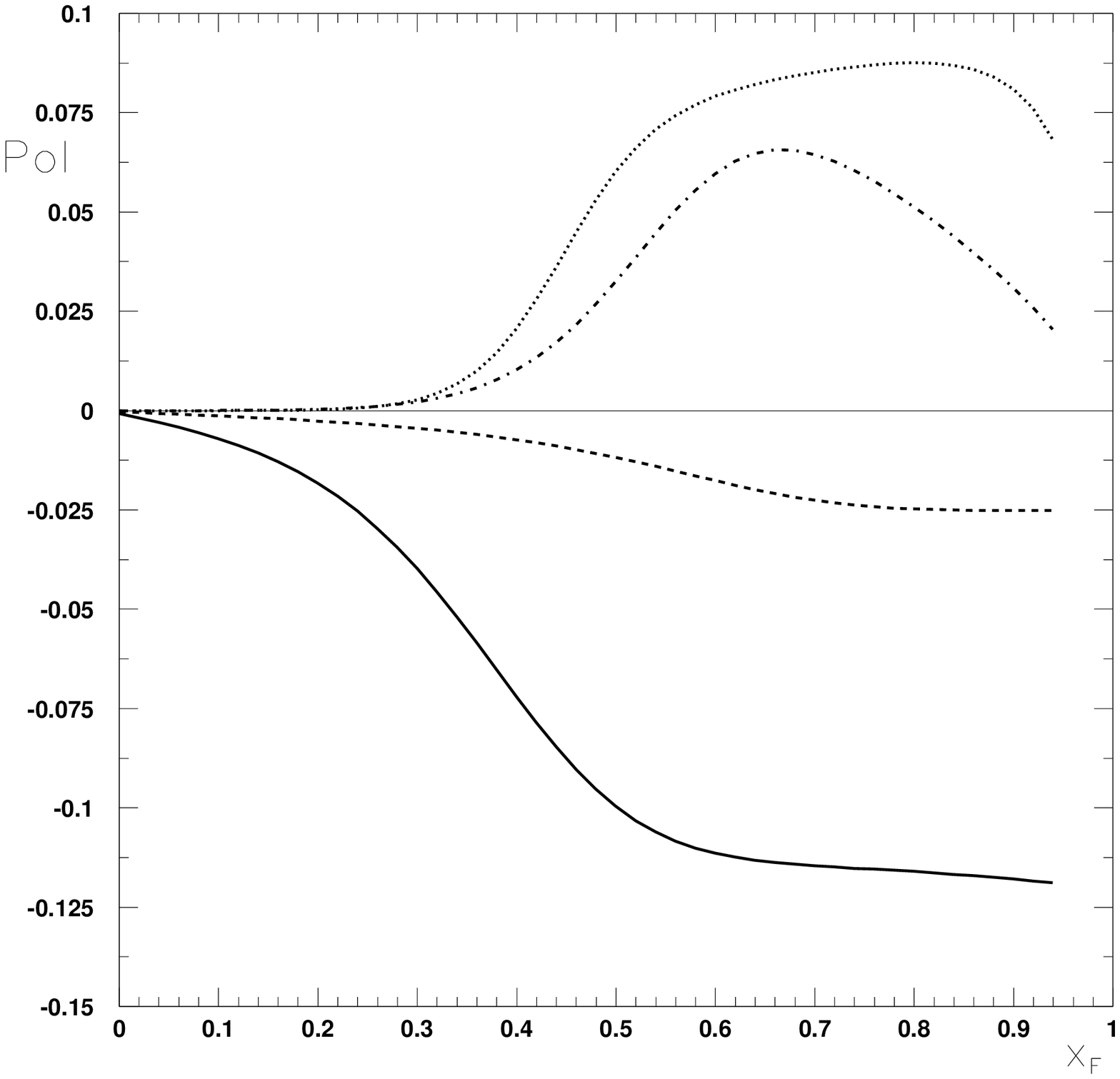,height=6.0in} \caption{} \label{fig-pol}
\end{figure}
\end{document}